# Genuine full characterization of partially coherence beam

Xingyuan Lu[1,††], Zhuoyi Wang[1,††], Qiwen Zhan[2,*], Yangjian Cai[3,†], Chengliang Zhao[1,‡]

[1]*School of Physical Science and Technology, Soochow University, Suzhou 215006, China.*
[2]*School of Optical-Electrical and Computer Engineering, University of Shanghai for Science and Technology, Shanghai 200093, China.*
[3]*Shandong Provincial Engineering and Technical Center of Light Manipulations & Shandong Provincial Key Laboratory of Optics and Photonic Device, School of Physics and Electronics, Shandong Normal University, Jinan 250358, China.*

For partially coherent light fields with random fluctuations, the intensity distributions and statistics have been proven to be more propagation robust compared with coherent light. However, its full potential in practical applications has not been realized due to the lack of four-dimensional optical field measurement. Here, a general modal decomposition method of partially coherent light field is proposed and demonstrated. The decomposed random modes can be used to, but not limited to, reconstruct average intensity, cross spectral density and orthogonal decomposition properties of the partially coherent light fields. Due to its versatility and flexibility, this method provides a powerful tool to further reveal light field invariant or retrieve embedded information after propagation through complex media. The Gaussian-shell-model beam and partially coherent Gaussian array are used as examples to demonstrate the reconstruction and even prediction of second-order statistical characteristics. This method is expected to pave the way for applications of partially coherent light in optical imaging, optical encryption and anti-turblence optical communication.

Invarience in intensity [1], coherence [2], and polarization [3], always attracts great interest, especially when a light beam passes through a complex media. However, different from the intensity, invariance in polarization and coherence is difficult to measure. Recently, with the benefit of basis-indepdent measurement sheme, the polarization inhomogeneity, which defines vectorial structured light, has been proven to be invariant even in turblence. However, for a randomly fluctuated partially coherent light field [4], the invariance in coherence strucures is only found for specially designed model [2]. For general partially coherent light, the research into its propagation invariance has been hindered by the lack of genuine four-dimensional optical field measurement scheme.

The four-dimensional cross-spectral density (CSD) function in the space-frequency domain, that is, the statistical average of the fluctuating electric fields [5], has been proposed for characterizing the partially coherent beam. Due to the unique advantages in obtaining high signal-to-noise ratio [6-8], high-resolution imaging [9], and anti-turblence transmission [10] of partially coherent light, plenty of schemes for generating partially coherent light have been proposed, such as those based on Van Cittert–Zernike theorem [11], modal superposition [12-13], rotating difusers, light emitting diodes, lithium niobate [14] and disorder-engineered statistical photonic platform [15].

To characterize a partially coherent light field, Young's double slit interference experiment has been wildly used [15]. However, the statistical characteristics of complex partially coherent light cannot be fully verified by two-point interference. Thus, various methods have been proposed to measure the complex-valued CSD distribution, such as generalized Hanbury Brown-Twiss experiment [16, 17], diffraction through obstacles [18], and holography [19, 20]. Although modal superposition has been widely used in generating partially coherent beam [11-15], exisiting methods generally focus on measuring the coherence of point to point or surface. Modal decomposition, on the other hand, has never been realized experimentally for partially coherent light. Even the best measurement of the four-dimensional partially coherent light field can only present as many features as possible on certain fixed plane [19, 20], which cannot be regarded as genuine full characterization of the partially coherent light field.

In this work, a universal experimental modal decomposition scheme is proposed and demonstrated. The flexibility of characterizing the partially coherent beam with decomposed random modes is shown with a Gaussian-shell-model (GSM) beam and a partially coherent Gaussian array beam. The second-order statistics, including but not limited to average intensity and CSD, can be freely reconstructed with decomposed modes. No reference arm is introduced, and no prior knowledge of light field is required. More importantly, the second-order statistic on the source plane can be retrieved with angular spectrum propagation, as long as the transmission matrix is known or can be measured beforehand and the system is linear and time invatient. The interference properties predicted with the decomposed modes are highly consistent with the experimentally captured ones. This work has potential applications of measuring the second-order statistics properties [21], coherence manipulation [22], optical imaging [23, 24], topological charge measurement [25, 26], and optical information encryption [2, 27].

Consider a scalar partially coherent source that propagates closely along the $z$ axis. The second-order statistical properties of the source field can be characterized in the space-frequency domain by an electric CSD function $W(\mathbf{r}_1, \mathbf{r}_2) = \langle E^*(\mathbf{r}_1)E(\mathbf{r}_2) \rangle$, where the asterisk and angle brackets denote complex conjugate and ensemble average; $\mathbf{r}_1$ and $\mathbf{r}_2$ are two transverse position vectors; $E(r)$ denotes the field fluctuating in a direction perpendicular to the $z$ axis. If we can obtain the full information of CSD, the partially coherent beam can be back propagated to the source. On the other hand, the statistical properties of the partially coherent beam passing through optical system or complex media can be predicted with this knowledge as well.

Partially coherent beams can also be taken as an incoherent superposition of fully coherent beams [28]. Thus, in general, one can represent the CSD function of a statistically stationary partially coherent source with orthogonal decomposition, or random modes decomposition [12],

$$W(\mathbf{r}_1, \mathbf{r}_2) = \sum_{m=1}^{M} P_m^*(\mathbf{r}_1)P_m(\mathbf{r}_2). \quad (1)$$

Here, $\{P_m\}$ represents random modes and $M$ is the total number of modes. The CSD satisfies a pair of Helmholtz equations and it contains information about the spectral density (i. e. average intensity) $I(\mathbf{r}) = W(\mathbf{r}, \mathbf{r})$. Based on Eq. (1), the reconstruction of these random modes can help to reconstruct the complete CSD information of a partially coherent beam. In the experimental measurement, the ptychographic coherent diffractive imaging (CDI) [29] is introduced to realize the modal decomposition.

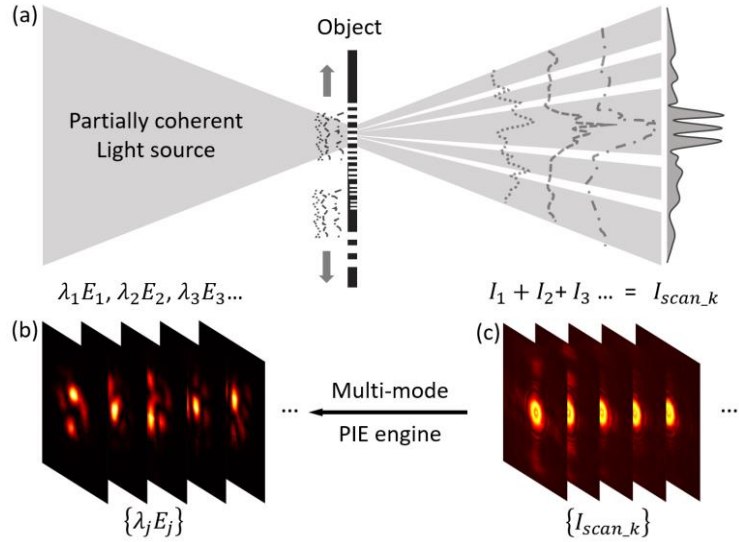

FIG. 1. Principle and diagram of modal decomposition for a partially coherent beam. (a) A partially coherent light composed of mixed modes $\{\lambda_j E_j\}$ illuminates an object and the averaged superposed diffraction patterns are detetected by the camera. A set of (b) superposed modes $\lambda_j E_j$ are reconstructed from (c) a set of superposed diffraction patterns $\{I_{scan\_k}\}$ with ptychography iterative engine (PIE) engine. Each diffraction pattern corresponds to an averaged intensity $I_{scan\_k}$ at each scan position.

Figure 1 shows the principle and diagram of the modal decomposition for a partially coherent beam. The partially coherent light composed of mixed modes $\{\lambda_j E_j\}$ illuminates an object and the averaged superposed diffraction patterns $\{I_{scan\_k}\}$ are detetected by the camera. $\lambda_j$ means the mode weigh of corresponding $E_j$ [12]. When the partially coherent beam $W(\mathbf{r}_1, \mathbf{r}_2)$ illuminates the object $O(r)$, the CSD of the far field can be approximated as

$$W(\boldsymbol{\rho}_1, \boldsymbol{\rho}_2) = \mathfrak{F}\{W(\mathbf{r}_1, \mathbf{r}_2)O^*(\mathbf{r}_1)O(\mathbf{r}_2)\}$$
$$= \mathfrak{F}\{\sum_{m=1}^{M} P_m^*(\mathbf{r}_1)O^*(\mathbf{r}_1) P_m(\mathbf{r}_2)O(\mathbf{r}_2)\}, (2)$$

where $\boldsymbol{\rho}$ is the coordinate on the detector plane and $\mathfrak{F}$ represents the Fourier transform operator. For other optical systems, Fourier transform operator can be replaced with other operators. The diffraction intensity $I(\boldsymbol{\rho})$ equals $W(\boldsymbol{\rho}, \boldsymbol{\rho})$. Then, using the multi-mode ptychography iterative engine (PIE), the object $O(r)$ and a set of probes $\{P_m(r)\}$ can be updated from a set of averaged diffraction intensities $\{I_{scan\_k}(\boldsymbol{\rho})\}$ [29]. The examples of GSM beam and partially coherent Gaussian array beam are illustrated in Figure 2 and Figure 3, respectively.

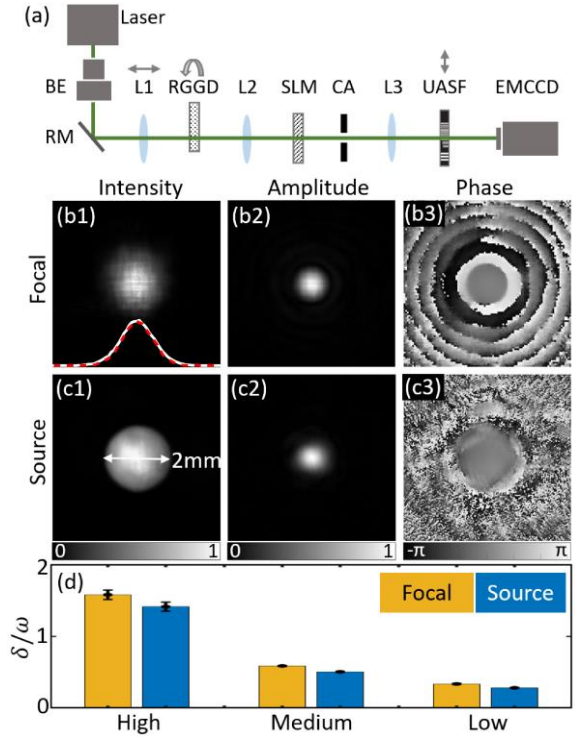

FIG. 2. Second-order statistics reconstruction of partially coherent GSM from experimentally decomposed modes. (a) Schematic diagram of the experimental setup. Focal average intensity, CSD amplitude and phase patterns of a GSM on the (b1-b3) focal and (c1-c3) source plane. Illustration in (b1) shows a fitting of actually captured intensity (white-solid curve) and reconstructed intensity (red-dashed curve). (d) shows the reconstructed ratio of beam waist and spatial coherent width for cases with high to low coherency. The error bars represent the standard deviation of ten measurement results.

A schematic diagram of the experimental setup is shown in Figure 2 (a). The partially coherent light source is generated with coherent laser beam and a rotating ground glass disk (RGGD) based on Van Cittert–Zernike theorem [11, 30]. The spatial coherence width $\delta_0$ of the collimated beam after lens L2 is determined by the focal spot size on the RGGD and the roughness of the RGGD together. The lens L1 can be shifted back and forth to realize the size control of focal spot. The larger the beam spot, the less coherent the generated partially coherent beam. A spatial light modulator (SLM) was used to modulate the amplitude profile of the output partially coherent beam. Then, a GSM or a partially coherent Gaussian array beam can be generated as the partially coherent light source shown in Figure 1(a). Here, BE is a beam expander; RM is a reflective mirror; CA is a circular aperture used to filter out the +1 diffraction order from SLM.

Taking the focal plane of lens L3 as measurement plane, an object (USAF) is placed perpendicular to optical axis, and the diffractive intensities are captured with a camera (EMCCD, iXon Life, Oxford). Based on the modal superposition theory [12], it can be taken as the object is illuminated by multiple electric field $P_m(\mathbf{r})$ and the resulted intensities superimpose on the camera plane. Here, the distance of USAF to the EMCCD is set as 146mm. In order to collect more diffraction information, overlap scanning is performed. The object was fixed on a two-dimensional mobile stage (CONEX-MFACC Newport), and 400 raw diffraction patterns were captured via 20 × 20 overlap scanning. The probe beam width is around 200um, and the step size is set as 40um. After 100 iterations, multiple probes can be reconstructed, and these probes are the final decomposed random modes.

To demosntrate the modal decomposition of GSM, a Gaussian amplitude mask with beam waist $\omega$ =1mm is loaded with an SLM. A hard-edged aperture function with 2mm diameter is also applied to limit the area to be reconstructed. From 20 × 20 diffraction patterns (see Figure 1(c)), intotal 64 random modes (examples shown in Figure 1(b)) are reconstructed. Figure 2(b1) shows the calculated focal average intensities by modal superposition, which gives $I(r) = W(r,r)$. In addition, the CSD $W(\mathbf{r}_1, \mathbf{r}_2)$ are calculated with Eq. (1), and an on-axis reference point ($\boldsymbol{r}_2 = [0, 0]$) is choosen to show the two-dimensional slice of CSD. The intensity, CSD amplitude and CSD phase shown in Figure 2(b1-b3) agree well with the theoretical results, that is Gaussian distribution. The fitting of actually captured focal intensity (white-solid curve) and reconstructed focal intensity (red-dashed curve) is illustrated in Figure 2(b1). The rings around the CSD amplitude central spot and the ring dislocations in the phase are caused by the hard-edge aperture on the source.

Focal lens system can be approximated as a linear and time invatient system and the transmission matrix is measurable. Thus, the random modes can be inversely propagated to the source plane via angular spectrum propagation. As shown in Figure 2(c1), the intensities calculated with inversely propagated random modes also agree very well with the ground truth. The boundary of the hard-edge aperture can be clearly observed, whose size is 2mm in width. Furthermore, the CSD amplitude distribution in the source plane also presents a Gaussian profile, as shown in Figure 2(c2), and the CSD phase shown in Figure 2(c3) is nearly uniform, because there is no phase term in the CSD function of GSM.

As the initial coherence width $\delta_0$ decreases, the focal and source intensities remain Gaussian profile, while the focal beam waist will increase gradually. Figure 2(d) shows the measured ratio of coherence/beam-waist (that is $\delta/\omega$) for different levels of coherence (high, medium and low). Ten datasets are analysed for each case and the height of each bar is set as the mean value of ten measurements. It shows that, the source ratio is almost the same as that on the focal plane, which is consistant with the theoretical prediction. The error bar shows the measurement error will decrease as the coherence decreases. The slight increase from source to focal plane is caused by the aperture applied on the source GSM. In addition, the uniform area in CSD phase will also shrink as the coherence decreases, because the reconstruction field of view is limited by the CSD amplitude profile, and the reconsruction ouside the profile will be noisy.

Rargarding the number of random modes used in iterative engine, when the coherence is reduced, the number of modes ($M$) should increase based on the oversampling theory of CDI [31]. In the GSM modal superposition simulation, when $M$ equals 25 for high coherence light source and $M$ equals 64 for lower coherence cases, the superimposed intensity and CSD matches well with that calculated with GSM Mathematica model.

To further evaluate the universality of the proposed measurement methods, the second-order statistics for a partially coherent Gaussian array beam [32] are measured, as shown in Figure 3. The Gaussian array mask is loaded with an SLM, and the modulated +1 order can be selected with CA or a 4$f$ lens system. The beam waist of each Gaussian spot is 0.3mm and the center-to-center distance is 0.8mm. Four array layouts are shown here. Each focused partially coherent Gaussian array beam illuminates the object plane, where the random modes are measured. Similar with previous calculation, the intensity and CSD are calculated with these random modes. It can be found that, the focal properties (Figure 3(a-d)) evolve into distributions different from the source plane (Figure 3(e-h)).

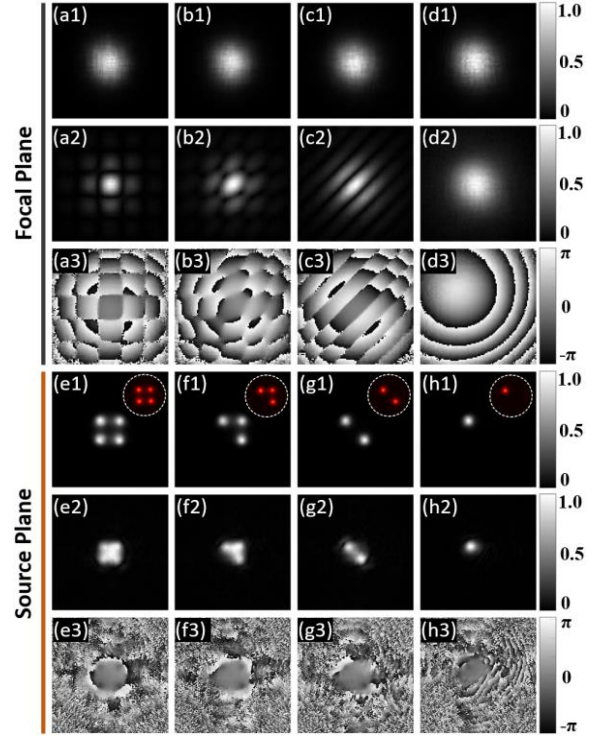

FIG. 3. Second-order statistics reconstruction of low coherence Gaussian array from experimentally decomposed modes. (a1-d1) Focal average intensity, (a2-d2) CSD amplitude and (a3-d3) CSD phase are reconstructed with random modes. (e1-h1) Source average intensity, (e2-h2) CSD amplitude and (e3-h3) CSD phase are reconstructed by inverse propagation of focal random modes to the source plane. Illustrations in (e1-h1) are the actually captured intensities on the source plane.

For high coherence light source, the focal pattern differs as the Gaussian array changes. However, as the coherence decreases, all these focal intensities tend to be Gaussian distributed, as shown in Figure 3(a1-d1). Different from GSM, the CSD of partially coherent Gaussian array is complex-valued, whose structure has more features. For the single Gaussian spot case, the focal intensity and CSD amplitude are similar with that shown in Figure 2, while the phase shows difference due to the off-center of Gaussian spot on the source plane. Furthermore, the intensity and CSD information on the source plane are also calculated by inverse propagation of focal random modes to the source plane. The actually captured intensities are shown in illustrations of Figure 3(e1-h1). Obviously, the Gaussian array structures are reconstructed very well, even the focal intensity has no more features to distinguish the array layout. This study may find applications in optical encrypted transmission.

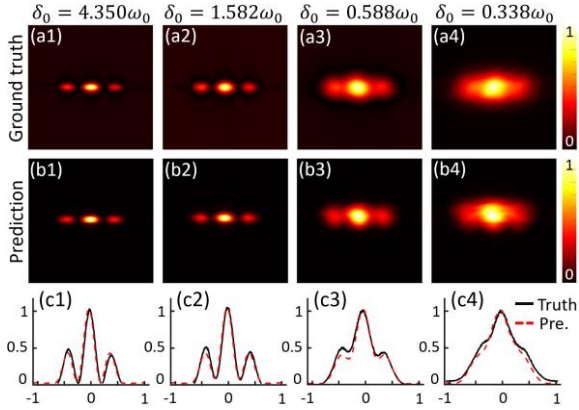

FIG. 4. Interference prediction of a GSM beam passing through a double-slit. (a1-a4) Experimentally captured interference fringes of a GSM beam passing through a double-slit under different spatial coherence. (b1-b4) Corresponding prediction results. (c1-) Cross lines of predicted intensity (red dotted lines) and real-captured intensity (black lines).

After reconstructing the source randome modes via inverse propagation of focal random modes, the interference properties of a GSM passing through a double slit are predicted. Slit width is 0.2 mm, and center-to-center distance is 0.4 mm. Figure 4(a1-a4) show the real-captured interference patterns (taken as ground truth) under different coherence widths and Figure 4(b1-b4) show corresponding predicted interference patterns. The one-dimensional intensity fitting curves are shown in Figure 4(c1-c4). Obviously, the predicted results agree well with the ground truth. More importantly, such capability opens completely new opportunities for the future applications that requires close loop iterative control of partially coherent light propagation.

To summarize, experimental modal decomposition of partially coherent beam have been demonstrated via scanning an object within the beam to be measured. This method provides a flexible and high-precision way to fully characterize the second-order statistical properties of randomly fluctuated partially coherence beams. This scheme integrates all the functions of the existing measurement schemes of partially coherent beam, while only a scattering object and a camera are needed. In addition, the decomposed modes can be inversely propagated to the source plane, or any other transverse planes of the linear and time invatient system, as long as the transmission matrix is known or measurable. Furthermore, the statistical properties of un-occuring diffraction or interference can also be predicted.

Especially for the low coherence beam whose intensity contains few information or the beam whose information is hidden in the source CSD distribution, the modal decomposition scheme demonstrated in this work is powerful and indispensable. It also has potentional applications in the characterization of the vector partially coherent beam with non-uniform CSD function or the superimposed beam with complex-valued screens. The idea of iterative reconstruction is also expected to be feasible in designing minimal-mode partially coherent light source in micro-nano system where lithium niobate crystal [33] or metasurface [15] is used.

This work is supported by the National Key Research and Development Program of China (No. 2022YFA1404800, No. 2019YFA0705000), National Natural Science Foundation of China (No. 12204340, No. 12174280, No. 12192254, No. 11974218, No. 92250304, No. 92050202), China Postdoctoral Science Foundation (2022M722325), Priority Academic Program Development of Jiangsu Higher Education Institutions, Tang Scholar, Local Science and Technology Development Project of the Central Government (No. YDZX20203700001766), Key Lab of Modern Optical Technologies of Jiangsu Province (KJS2138).

———————

††These authors contributed qually to this work
*qwzhan@usst.edu.cn
†yangjian_cai@163.com
‡zhaochengliang@suda.edu.cn